\documentclass[prd,aps,twocolumn,showpacs,preprintnumbers,superscriptaddress,floatfix,nofootinbib,10pt]{revtex4-1}

\usepackage{graphicx}
\usepackage{epsfig}
\usepackage{epstopdf}
\usepackage{hyperref}
\usepackage{amsmath}
\allowdisplaybreaks[1]
\usepackage{amsfonts}
\usepackage{bm,amssymb}
\usepackage{multirow}
\usepackage{makecell}
\usepackage[dvipsnames]{xcolor}
\usepackage[normalem]{ulem}
\usepackage{longtable}
\usepackage{array}
\usepackage{subfigure}
\usepackage{booktabs}
\RequirePackage{microtype}

\newcommand{\Slash}[1]{\ooalign{\hfil/\hfil\crcr$#1$}}

\newcommand{\ucas}{\affiliation{School of Physical Sciences, University of Chinese Academy of Sciences, Beijing 100049, China}}
\newcommand{\itp}{\affiliation{CAS Key Laboratory of Theoretical Physics, Institute of Theoretical Physics,\\ Chinese Academy of Sciences, Beijing 100190, China}}

\begin{document}

\title{Exploring possible triangle singularities in the $\Xi^-_{b} \to K^- J/\psi \Lambda$ decay}

\author{Chao-Wei Shen} \email{shencw@ucas.ac.cn}
\ucas

\author{Hao-Jie Jing} \email{jinghaojie@itp.ac.cn}
\itp\ucas

\author{Feng-Kun Guo} \email{fkguo@itp.ac.cn}
\itp\ucas

\author{Jia-Jun Wu} \email{wujiajun@ucas.ac.cn}
\ucas

\date{\today}

\begin{abstract}

We analyze possible singularities in the $J/\psi \Lambda$ invariant mass distribution of the $\Xi^-_{b}~\to~K^- J/\psi \Lambda$ process via triangle loop diagrams. Triangle singularities in the physical region are found in 18 different triangle loop diagrams. Among those with $\Xi^*$-charmonium-$\Lambda$ intermediate states, the one from the $\chi_{c1} \Xi(2120) \Lambda$ loop, which is located around 4628~MeV, is found the most likely to cause observable effects. One needs $S$- and $P$-waves in $\chi_{c1} \Lambda$ and $J/\psi \Lambda$ systems, respectively, when the quantum numbers of these systems are $1/2^+$ or $3/2^+$. When the quantum numbers of the $\Xi(2120)$ are $J^P=1/2^+$, $1/2^-$ or $3/2^+$, the peak structure should be sharper than the other $J^P$ choices. This suggests that although the whole strength is unknown, we should pay attention to the contributions from the $\Xi^*$-charmonium-$\Lambda$ triangle diagram if structures are observed in the $J/\psi \Lambda$ invariant mass spectrum experimentally. In addition, a few triangle diagrams with the $D_{s1}^*(2700)$ as one of the intermediate particles can also produce singularities in the $J/\psi\Lambda$ distribution, but at higher energies above 4.9~GeV.

\end{abstract}

\maketitle

\section{Introduction}

In the earliest prediction of the hidden-charm baryon $N_{c\bar c}$ states above 4~GeV~\cite{Wu:2010jy}, which are now known as the $P_c$ states and have been reported by the LHCb Collaboration~\cite{Aaij:2015tga,Aaij:2019vzc}, the~$\Lambda_{c\bar c}$ states are also predicted with different masses and spin-parities.
Ten $\Lambda_{c\bar c}$ states are obtained in a coupled-channel model in Ref.~\cite{Xiao:2019gjd} and their couplings to $J/\psi \Lambda$ are found to be sufficiently large.
Similar to the discovery of the $P_c$ states in the  $J/\psi p$ channel~\cite{Xiao:2013yca}, the~$\Lambda_{c\bar c}$ states could be sought in the $J/\psi\Lambda$ invariant mass distributions.
In Ref.~\cite{Wang:2019nvm}, ten isoscalar molecular pentaquarks in the $\bar{D}^{(*)} \Xi_c$, $\bar{D}^{(*)} \Xi_c^{\prime}$ and $\bar{D}^{(*)} \Xi_c^*$ systems are predicted by calculating the effective potentials with the chiral effective field theory up to the next-to-leading order.
These hidden-charm pentaquark states with strangeness are investigated in the quark model including the color octet type and color singlet type~\cite{Irie:2017qai} and the one-boson exchange model~\cite{Chen:2016ryt}.
The decay behaviors of the $\Lambda_{c \bar c}$ states in various hadronic molecular assumptions with different spin-parities are studied in Ref.~\cite{Shen:2019evi} and these decay patterns are suggested to help understanding the internal structures after the experimental searches.
Although there are various predictions for such strangeness baryons within hidden charm theoretically, there is no experimental evidence for them until~now.

Since the first prediction of the $\Lambda_{c\bar c}$ states, there arises various suggestions of searching for these states experimentally.
The decay of $\Lambda_{b}$ into $J/\psi K^{0} \Lambda$ states is studied in Ref.~\cite{Lu:2016roh} and a clear peak is predicted in the $J/\psi \Lambda$ invariant mass distribution.
In Ref.~\cite{Feijoo:2015kts}, the~authors propose the $\Lambda _b \to J/\psi \eta \Lambda$ decay and find that a resonance signal with similar strength as the case of $P_c$ should be observed in the $J/\psi \Lambda$ invariant mass spectrum.
The $\Xi_b^{-} \to J/\psi K^{-} \Lambda$ reaction is discussed in Ref.~\cite{Chen:2015sxa} and a hidden-charm pentaquark state with strangeness is expected to be clearly seen.
In the last process, the~$K^-$ meson in the final state is a charged meson, and~it makes this process suitable for experimental measurement.
Actually, the~possible peak structures in the $J/\psi \Lambda$ invariant mass spectrum in the $\Xi_b^{-} \to J/\psi K^{-} \Lambda$ decay have already been sought by the LHCb Collaboration~\cite{Aaij:2017bef}, while the results of these data should still be under analysis.
At this point, it is worthwhile to mention that the $\Lambda_b^0\to J/\psi\Lambda \pi^+\pi^-$ should also be a promising process to seek for the $\Lambda_{c\bar c}$ states.
In this paper, we will study the $\Xi_b^{-} \to K^{-} J/\psi \Lambda$ decay and focus on another mechanism, triangle singularity, which could also produce peak structures in the $J/\psi \Lambda$ invariant mass~spectrum.

A triangle singularity is due to the simultaneous on-shellness of three intermediate particles in a loop diagram, and~at the same time the interactions at all vertices can happen classically~\cite{Coleman:1965xm}. It is a logarithmic singularity, and~is able to produce peaks in energy distributions. It has been known for a long time~\cite{Karplus:1958zz,Landau:1959fi,Bjorken:1959fd}, and~there were debates in the 1960s on whether it could produce observable effects~\cite{Schmid:1967ojm}.
Nevertheless, it was difficult to find reactions that satisfy the rigorous kinematic conditions and at the same time can evade the subtle interfering cancellation between the tree-level and triangle diagram, dictated by the Schmid theorem at that time~\cite{Schmid:1967ojm}.
\textls[-15]{In the last 10 years or so, suggestions were proposed that many reactions which can be (or have been) measured at modern high-energy experiments could match the kinematic conditions and might receive important contributions from triangle singularities; then the study of triangle singularity started to bring a lot interest until now~\cite{Wu:2011yx, Aceti:2012dj, Wu:2012pg, Wang:2013cya,Wang:2013hga, Ketzer:2015tqa, Achasov:2015uua, Liu:2015taa,
Liu:2015fea, Guo:2015umn, Szczepaniak:2015eza, Guo:2016bkl, Bayar:2016ftu, Wang:2016dtb, Pilloni:2016obd, Xie:2016lvs, Szczepaniak:2015hya, Roca:2017bvy,
Debastiani:2017dlz, Samart:2017scf, Sakai:2017hpg, Pavao:2017kcr, Xie:2017mbe, Bayar:2017svj,
Liang:2017ijf, Oset:2018zgc, Dai:2018hqb, Dai:2018rra, Guo:2019qcn, Liang:2019jtr, Nakamura:2019emd, Du:2019idk,
Liu:2019dqc, Jing:2019cbw, Braaten:2019gfj, Aaij:2019vzc, Sakai:2020ucu, Sakai:2020fjh, Molina:2020kyu, Braaten:2020iye, Alexeev:2020lvq, Ortega:2020ayw}.~For a recent review, we refer to Ref.~\cite{Guo:2019twa} (see Tables~\ref{table:TS} and \ref{table:TSbc}
 therein for reactions proposed since the 1990s)}.

The fact that triangle singularities may affect the extraction of the $P_c$ resonance parameters from the $\Lambda_b \to J/\psi K^- p$ reaction was immediately noticed~\cite{Guo:2015umn,Liu:2015fea}.
Later, the~$\Lambda(1890)\chi_{c1}p$ triangle was identified to be the most special from an analysis of many triangle diagrams with various possible combinations of a a charmonium, a~proton and a $\Lambda^*$~\cite{Bayar:2016ftu}.
In this case, both triangle singularity and threshold effect occur at about 4.45~GeV as already pointed out in Ref.~\cite{Guo:2015umn}.
Once the $\chi_{c1} p$ is in an $S$-wave, such that the quantum numbers of the $J/\psi p$ could be $J^P=1/2^+$ or $3/2^+$, it would generate a narrow peak in the $J/\psi p$ invariant mass distribution.
The updated LHCb measurement~\cite{Aaij:2019vzc} does not have a peak at 4.45~GeV, but~there is a dip at this energy between the two narrow $P_c(4440)$ and $P_c(4457)$ peaks. Thus, the~possibility of interference of such a triangle singularity with pentaquark signals is still there, and~might be partly responsible for the special production properties of the $P_c$ states~\cite{Sakai:2019qph}.
However, for~the cases that the spin and parity of the $P_c$ are $3/2^-$ or $5/2^+$, as~considered in, e.g.,~Refs.~\cite{Chen:2015moa, Roca:2015dva, He:2015cea, Lin:2017mtz, He:2016pfa,
Chen:2016otp, Xiang:2017byz}, the~triangle singularity contribution from the $\Lambda^*(1890)\chi_{c1}p$ loop would be severely suppressed.
In the present work, we will investigate possible triangle singularities in the $\Xi_b^{-} \to J/\psi K^{-} \Lambda$ reaction in a similar way to Ref.~\cite{Bayar:2016ftu}.

This work is organized as follows.
After the introduction, we first show the theoretical framework and an analysis of possible triangle singularities in Section~\ref{sec:theory}, and~the corresponding discussions are presented.
Then in Section~\ref{sec:analysis}, a~detailed analysis and discussions of the amplitude with $\Xi(2120)\chi_{c1}\Lambda$ are given, which is followed by a brief summary in Section~\ref{sec:summary}.

\begin{table}[tb]
	\caption{\label{table:TS}The charmonium and $\Xi^{* -}$ states in Fig.~\ref{Fig:1stdiag} that could generate triangle singularities and the positions of the corresponding singularities.
	Among the listed hyperons, $\Xi(2030)$ is a three-star state; $\Xi(2250)$ and $\Xi(2370)$ are two-star states; $\Xi(2120)$ and $\Xi(2500)$ are one-star states~\cite{Zyla:2020}.
	}
	\renewcommand\arraystretch{1.5}
\begin{tabular}{p{1.0cm}<{\centering}p{2.0cm}<{\centering}p{2.0cm}<{\centering}p{3.0cm}<{\centering}}
\hline
No. & $c \bar c$ & $\Xi^{* -}$ & Position of triangle singularity (MeV) \\
\hline
1 	& $J/\psi$ 			& $\Xi(2500)$ 	& 4232 \\
2 	& $\chi_{c0}$ 	& $\Xi(2250)$ 	& 4546 \\
3 	& $\chi_{c0}$ 	& $\Xi(2370)$ 	& 4665 \\
4 	& $\chi_{c1}$ 	& $\Xi(2120)$ 	& 4628 \\
5 	& $\chi_{c1}$ 	& $\Xi(2250)$ 	& 4696 \\
6 	& $\chi_{c2}$ 	& $\Xi(2120)$ 	& 4680 \\
7 	& $h_c$ 				& $\Xi(2120)$ 	& 4644 \\
8 	& $h_c$ 				& $\Xi(2250)$ 	& 4730 \\
9 	& $\eta_c(2S)$ 	& $\Xi(2030)$ 	& 4754 \\
10 	& $\eta_c(2S)$ 	& $\Xi(2120)$ 	& 4797 \\
11 	& $\psi(2S)$ 		& $\Xi(2030)$ 	& 4810 \\
\hline
\end{tabular}
\end{table}
\unskip

\begin{table}[tb]
	\caption{\label{table:TSbc}The $D_s^{* -}$, $\Xi_c^{* 0}$ and intermediate particle 3 in Fig.~\ref{Fig:2nddiag} and Fig.~\ref{Fig:3rddiag} that could generate triangle singularities and the positions of the corresponding singularities.
	}
	\renewcommand\arraystretch{1.5}
\begin{tabular}{p{0.3cm}<{\centering}p{1.9cm}<{\centering}p{1.6cm}<{\centering}p{1.7cm}<{\centering}p{2.3cm}<{\centering}}
\hline
No. & $D_s^{* -}$ & $\Xi_c^{* 0}$ &	Intermediate particle 3	& Position of triangle singularity (MeV) \\
\hline
1 	& $\bar D_{s1/s3}^*(2860)$ & $\Xi_c(2930)$ & $\bar{D}^0$ 		& 4838 \\
2 	& $\bar D_{s1}^*(2700)$ 		& $\Xi_c(3055)$ & $\bar{D}^0$ 		& 4922 \\
3 	& $\bar D_{s1}^*(2700)$	 	& $\Xi_c(3080)$ & $\bar{D}^0$ 		& 4957 \\
4 	& $\bar D_{s1/s3}^*(2860)$ & $\Xi_c(2930)$ & $\bar{D}^{* 0}$ & 4959 \\
5 	& $\bar D_{s1}^*(2700)$ 		& $\Xi_c(3080)$ & $\bar{D}^{* 0}$ & 5089 \\
6 	& $\bar D_{s1}^*(2700)$		& $\Xi_c(3080)$ & $\Lambda_c^+$ 	& 4990 \\
7 	& $\bar D_{s1/s3}^*(2860)$	& $\Xi_c(2930)$	& $\Lambda_c^+$ 	& 5149 \\
\hline
\end{tabular}
\end{table}

\section{Theoretical Framework and Analysis Results of All Possible Triangle~Singularities} \label{sec:theory}

In the process of a triangle loop diagram, the~initial particle first decays to two intermediate particles $A$ and $B$.
Then particle $A$ continues to decay into two particles, one of which is a final state and the other is the intermediate particle $C$.
This intermediate particle $C$ could interact with particle $B$ to generate final state particles.
The kinematic conditions for a triangle singularity in the physical region are~\cite{Schmid:1967ojm,Bayar:2016ftu}: (1) the three intermediate particles $A$, $B$ and $C$ are all on the mass shell at the same time; (2) the velocity of particle $C$ is larger than that of particle $B$ and is in the same direction in the rest frame of the initial state.
In other words, triangle singularity corresponds to the situation that the loop reaction has actually occurred as a classical process~\cite{Coleman:1965xm}, rather than a virtual process.
We are interested in the possible triangle singularity effects in the $J/\psi\Lambda$ invariant mass distribution. The~region of triangle singularity peak is limited to a small region starting from the threshold of two particles that rescatter into the final-state $J/\psi\Lambda$ (for a detailed analysis, see Ref.~\cite{Guo:2019twa}).

\begin{figure}[h]
\centering
\subfigure[]{\label{Fig:1stdiag}
	\includegraphics[width=0.5\linewidth]{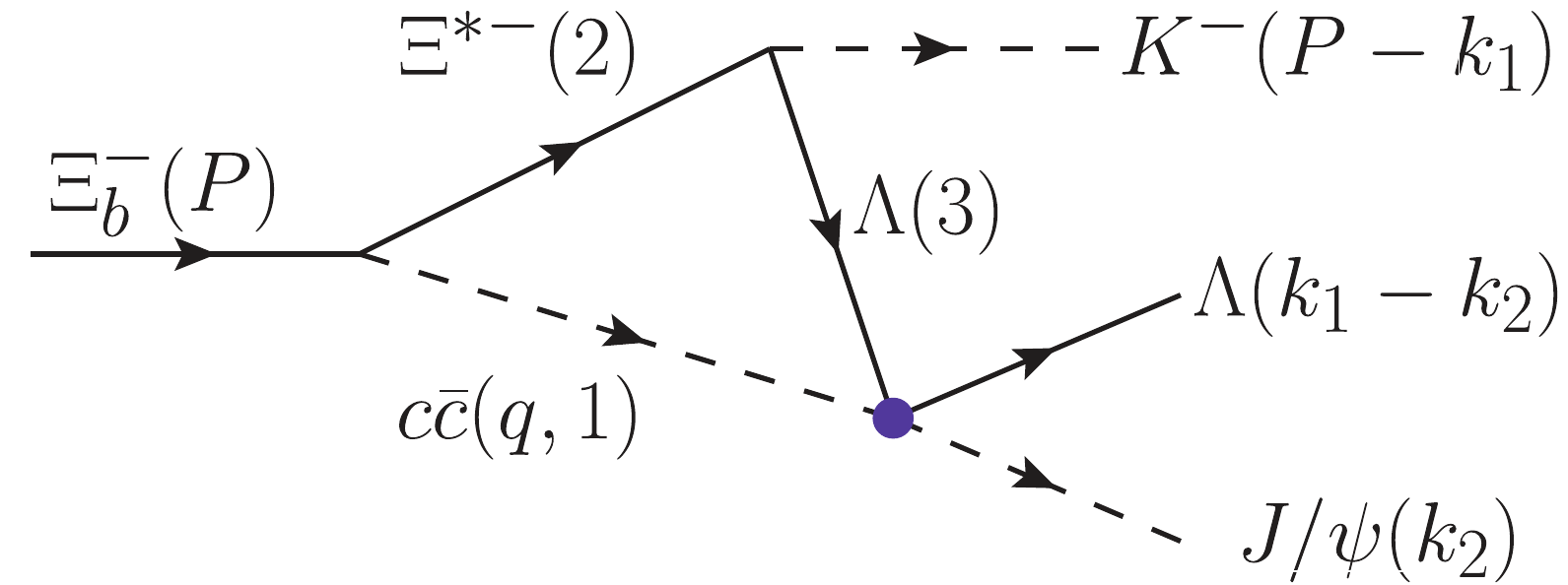}}
  \subfigure[]{\label{Fig:2nddiag}
	\includegraphics[width=0.45\linewidth]{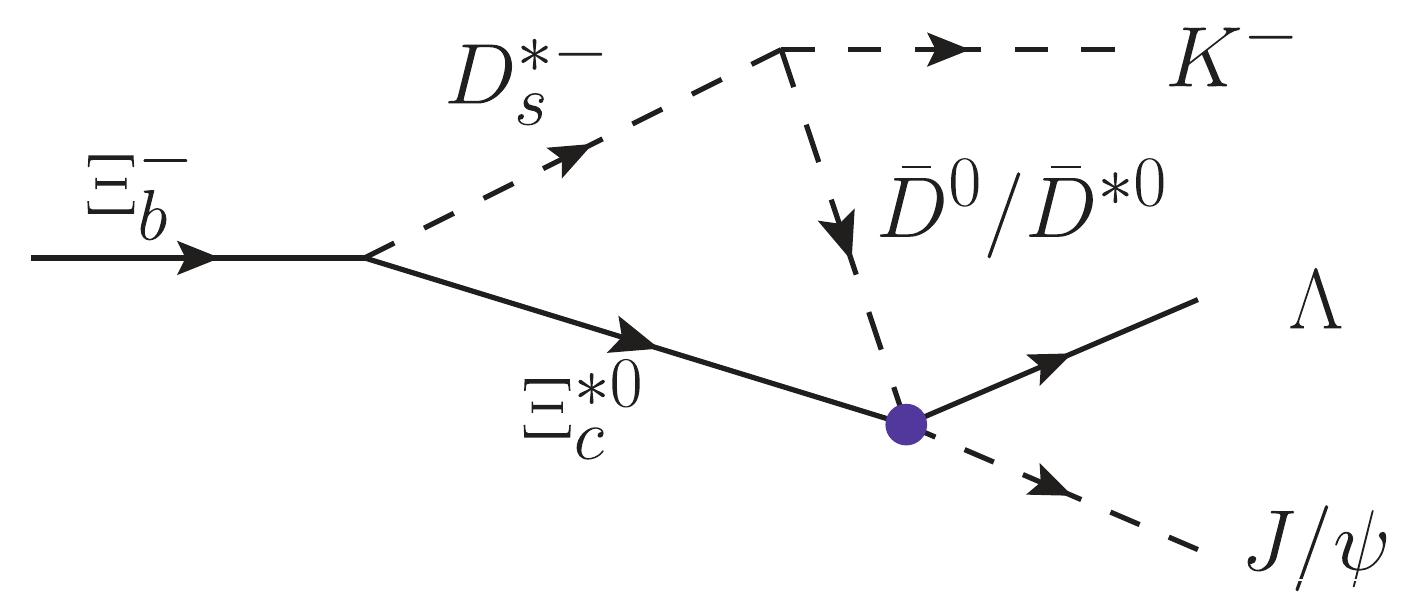}} \\ \ \
  \subfigure[]{\label{Fig:3rddiag}
	\includegraphics[width=0.45\linewidth]{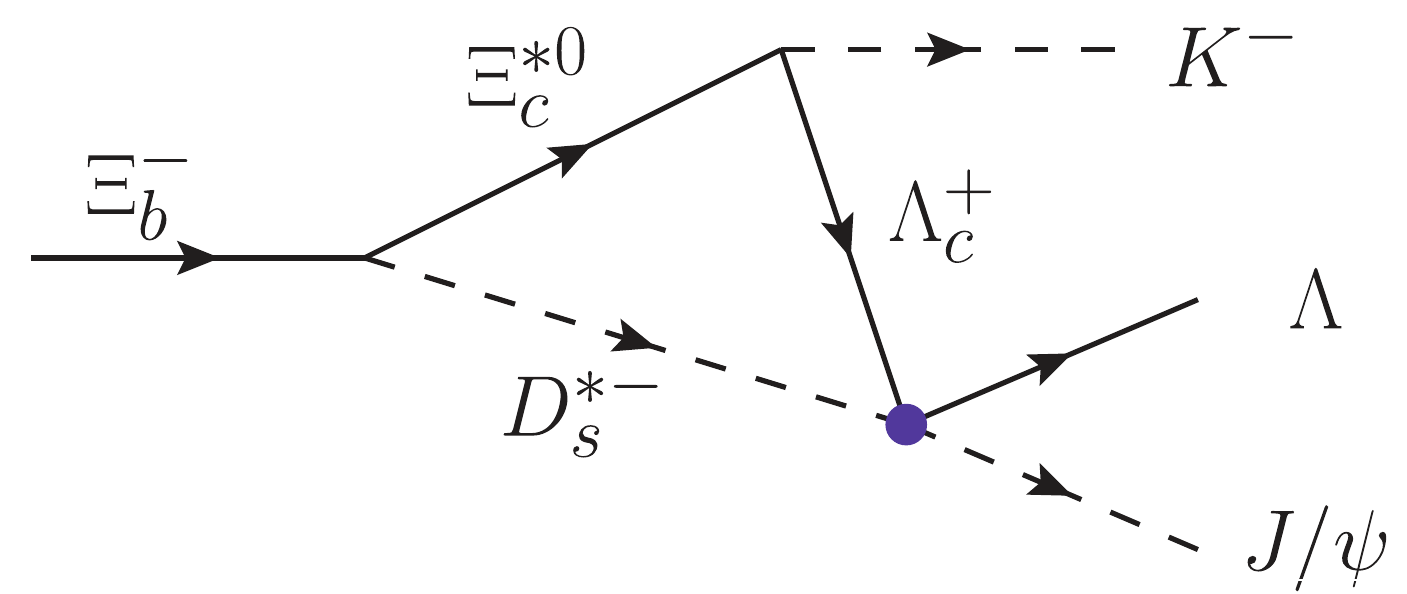}} \ \
  \subfigure[]{
	\includegraphics[width=0.45\linewidth]{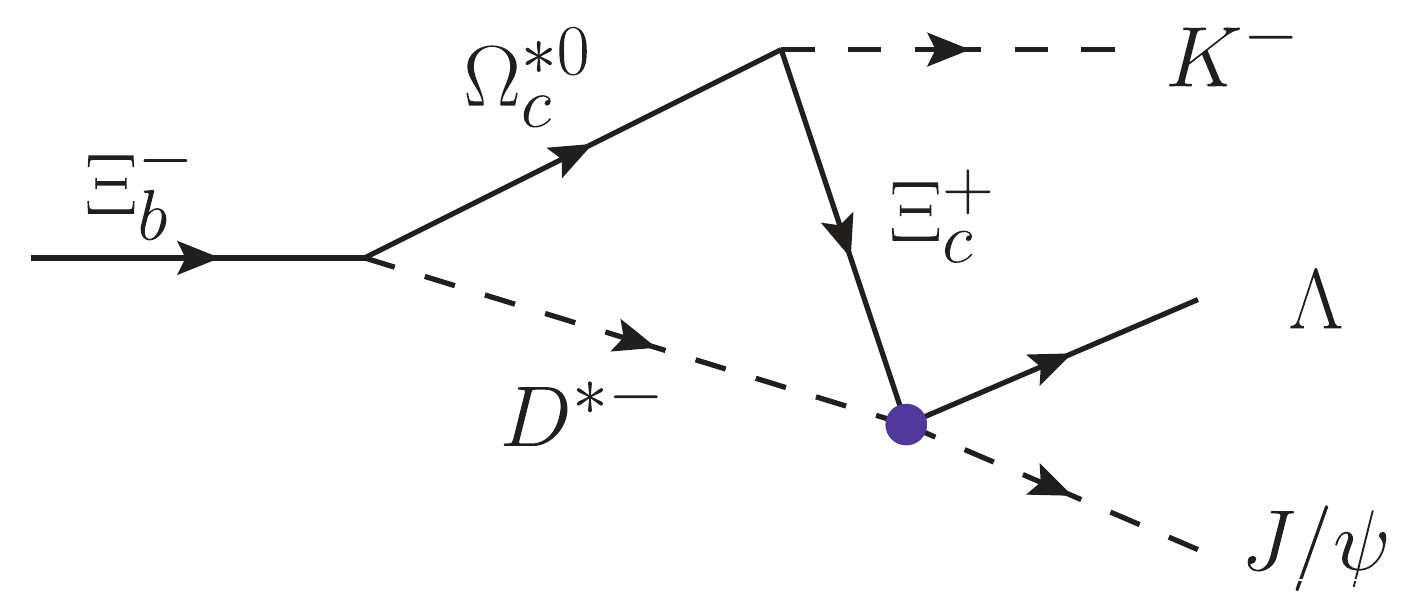}}
  \caption{Four possible kinds of triangle diagrams for the 
 $\Xi_b^- \to K^- \Lambda J/\psi$ decay that may have a triangle singularity of interest. The~momentum of each particle and the numbering of the exchanged particles used here are given in the brackets in the first diagram.
  \label{Fig:feyndiag4}}
\end{figure}

There are four possible kinds of triangle diagrams for the $\Xi_b^- \to K^- \Lambda J/\psi$ decay that may have a triangle singularity in the physical region (when the widths of the intermediate particles are neglected) as shown in Figure~\ref{Fig:feyndiag4}.
Using the intuitive equation derived in Ref.~\cite{Bayar:2016ftu}, we find that the physical-region triangle singularity emerges in the process of Figure~\ref{Fig:feyndiag4}a--c.

The $\Xi^{* -}$ in Figure~\ref{Fig:feyndiag4}a represents all the ten $\Xi$ resonances in the Review of Particle Physics (RPP)~\cite{Zyla:2020}, including $\Xi(1530)$, $\Xi(1620)$, $\Xi(1690)$, $\Xi(1820)$, $\Xi(1950)$, $\Xi(2030)$, $\Xi(2120)$, $\Xi(2250)$, $\Xi(2370)$ and $\Xi(2500)$.
For the charmonium states $c \bar c$, we take $\eta_c$, $J/\psi$, $\chi_{c 0}$, $\chi_{c 1}$, $\chi_{c 2}$, $h_c$, $\eta_c(2S)$ and $\psi(2S)$ into consideration.
We take all the listed $\Xi_c^{* 0}$ and $D_s^{* -}$ in RPP~\cite{Zyla:2020} into account for the exchanged particles in Figure~\ref{Fig:feyndiag4}b,c.

Through checking the two conditions of triangle singularity for all possible combinations of the considered intermediate particles mentioned above, we find that there are in total 11 cases for Figure~\ref{Fig:feyndiag4}a and 7 cases for Figure~\ref{Fig:feyndiag4}b,c that would have singularities in the physical region.
The charmonium and $\Xi^{* -}$ states in Figure~\ref{Fig:feyndiag4}a, which could generate peak structures in the $J/\psi \Lambda$ mass spectrum, and~the corresponding positions of these triangle singularities are given in Table~\ref{table:TS}.
In Table~\ref{table:TSbc}, we present the three exchanged particles of Figure~\ref{Fig:feyndiag4}b,c together with the corresponding triangle singularity positions,
which are generally higher than those from Figure~\ref{Fig:feyndiag4}a listed in Table~\ref{table:TS}.
Note that there is another combination, $\bar D_{sJ}(3040)\Xi_c(3645)\bar D^0$, which can give rise to a triangle singularity at 4511~MeV. However, so far the $D_{sJ}(3040)$ has only been observed in $D^*K$~\cite{Zyla:2020}, suggesting that it might have an unnatural parity and thus could not decay into $\bar D K$. Consequently, this combination is not listed in the table.
One~notices that both the $D_{s1}^*(2700)$ was observed in $B\to D\bar D K$~\cite{Brodzicka:2007aa,Lees:2014abp} and the $D_{s1/s3}^*(2860)$ were seen in the decays of $B_s\to  \bar D \bar K \pi$~\cite{Aaij:2014xza}. Similarly, the~$D_{s1}^*(2700)$ should be easily produced from the decays of the $\Xi_b$ associated with a $\Xi_c$ baryon, as~shown in the first vertex in Figure~\ref{Fig:feyndiag4}b,c.
Thus, among~the combinations listed in Table~\ref{table:TSbc}, the~ones with the $D_{s1}^*(2700)$ are expected to be more important than the others. Once a structure is observed in the $J/\psi\Lambda$ spectrum above 4.9~GeV in a future experiment, such~triangle singularities need to be considered.
In the rest of this paper, we will focus on the singularities in Table~\ref{table:TS}, whose situation is more~involved.

Since the triangle singularity is caused by the on-shellness of the intermediate particles, we first consider the scalar loop integral as follows,
\begin{eqnarray}
  I&=&i \int \frac{d^4 q}{(2\pi)^4} \frac1{(q^2-m_1^2+i m_1\Gamma_1)[(P-q)^2-m_2^2+i m_2\Gamma_2]} \nonumber\\
  &&\times \frac1{[(k_1-q)^2-m_3^2+i \epsilon]},
  \label{Eq:I}
\end{eqnarray}

\noindent where the momenta and the numbering of particle masses are as given in the first diagram in Figure~\ref{Fig:feyndiag4}, and~$\Gamma_1$ and $\Gamma_2$ are the widths of intermediate particles 1 and 2, respectively.
We can use $|I|^2$ to analyze the singular behavior of the diagrams in Figure~\ref{Fig:feyndiag4}.
It should be noted that the interactions of all the related vertices are not included here.
We should consider these interactions from a physical point of view to draw further~conclusions.

Let us turn to the three vertices in Figure~\ref{Fig:feyndiag4}a.
For the first vertex $\Xi_b \to \Xi^{*} c \bar c$, it is a weak decay and involves the Cabibbo--Kobayashi--Maskawa (CKM) matrix elements ${V_{bc}^*V_{cs}}$. The~strength should be comparable with that of $B \to K c \bar c$ for the same charmonium.
The branching ratios for different charmonia in the latter process, which can be found in the RPP~\cite{Zyla:2020}, are given in Table~\ref{table:1vertex}.
It can be seen that for the $\chi_{c2}$ and $h_c$, the~branching ratios are one-order-of-magnitude smaller than that of the other charmonia.
This indicates that the cases of the charmonium being the $\chi_{c2}$ or $h_c$ would not provide a strong signal of triangle singularity, and~can be safely neglected.
For $\chi_{c0}$, the~decay rates from the neutral and charged $B$ mesons are very different.
Both the two branching ratios of the $\chi_{c0}$ are smaller than the rest, although~in the charged case it is of the same order of magnitude.
We conclude that, for~the $\Xi_b \to \Xi^* c\bar{c}$ decay, the~cases where the intermediate charmonium is one of $J/\psi$, $\chi_{c1}$, $\eta_c(2S)$ and $\psi(2S)$ are expected to be the main decay~channels.

\begin{table}[tb]
	\caption{\label{table:1vertex}The branching ratios for different charmonia in $B \to K c \bar c$.}
	\renewcommand\arraystretch{1.5}
	\begin{tabular}{p{2.0cm}<{\centering}p{3.0cm}<{\centering}p{3.0cm}<{\centering}}
		\hline
		$c \bar c$ & $BR(B^0 \to c \bar{c} K^0)$  & $BR(B^+ \to c \bar{c} K^+)$ \\
		\hline
		$J/\psi$ & $8.73 \times 10^{-4}$ & $1.01 \times 10^{-3}$	\\
		$\chi_{c0}$ & $1.11 \times 10^{-6}$ & $1.49 \times 10^{-4}$	\\
		$\chi_{c1}$ & $3.93 \times 10^{-4}$ & $4.84 \times 10^{-4}$	\\
		$\chi_{c2}$ & $<1.5 \times 10^{-5}$ & $1.1 \times 10^{-5}$	\\
		$h_c$ &  & $<3.8 \times 10^{-5}$	\\
		$\eta_c(2S)$ &  & $4.4 \times 10^{-4}$	\\
		$\psi(2S)$ & $5.8 \times 10^{-4}$ & $6.21 \times 10^{-4}$	\\
		\hline
	\end{tabular}
\end{table}

For the vertex of $c \bar c \Lambda \to J/\psi \Lambda$, the~coupling strength is affected from the spin interactions between the charm quark and anticharm quark.
Once the $c \bar c$ has a spin flip, this vertex will be much suppressed due to breaking the heavy quark spin symmetry.
The quantum numbers of the final state $J/\psi$ are $1^-$, and~the spins of charm and anticharm quarks are in the same direction.
However, in~the charmonia $h_c$ and $\eta_c(2S)$, the~total spin of charm and anticharm quarks is $0$.
Therefore, the~$h_c \Lambda$ or $\eta_c(2S) \Lambda \to J/\psi \Lambda$ process would be suppressed because of the spin flip of the charm and anticharm quarks.
For the other charmonia considered here, the~couplings of $c \bar c \Lambda \to J/\psi \Lambda$ could be of the same order of magnitude but are still not determined. We will write the amplitudes with different $J^P$ assumptions that are subject to the Lorentz covariant orbital-spin coupling scheme~\cite{Zou:2002yy} to study their behavior in the next~section.

For the $\Xi^* \bar{K} \Lambda$ vertex, we can infer its strength from the $\Xi^* \to \Lambda \bar K$ process for each considered $\Xi^*$ state.
For the $\Xi(2030)$, the~$\Lambda \bar K$ state occupies about 20\% of all possible decay rates.
In the RPP, the~$\Lambda \bar K$ channel is the only seen and listed decay mode for the $\Xi(2120)$~\cite{Zyla:2020}, although~its exact ratio has not been measured yet.
The $\Lambda \bar K$ channel is not listed in the decay modes of the $\Xi(2250)$ and $\Xi(2370)$, and~actually all the decay channels for the former are three-body states, meaning that the $\Lambda \bar K$ mode might be suppressed for these two states.
For the $\Xi(2500)$, the~$\Lambda \bar K$ mode is listed, with~the ratio of branching fractions $\Gamma(\Lambda \bar K)/\left(\Gamma(\Xi\pi)+\Gamma(\Lambda \bar K)+\Gamma(\Sigma \bar K)+ \Gamma(\Xi(1530)\pi)\right) =  0.5\pm0.2$ measured in Ref.~\cite{Alitti:1969rb}, but~it is not clear whether the peak observed therein corresponds to the $\Xi(2500)$ as commented in RPP~\cite{Zyla:2020}.

Through the analysis of the three vertices above, for~the 11 cases listed in Table~\ref{table:TS}, nos.~2 and 3 and nos.~5--10 are suppressed; only nos.~1, 4 and 11 are left.
Now let us turn to the character of the intermediate particles in the loop, especially their quantum numbers, since they actually affect
a lot not only the magnitude but also the line shape of the singularity.
The quantum numbers are quite important as they directly determine the partial waves of the interaction, and~high partial wave interactions should be suppressed in the near-threshold region.
For the $\Xi(2030)$ state, its spin is equal to or greater than $5/2$~\cite{Hemingway:1977uw}.
It leads to high partial-wave ($l \ge 2$) interaction at the $\Xi(2030) \bar{K} \Lambda$ vertex, and~will not be considered further.
{We will discuss that how the high partial-wave interactions in the triangle loop would suppress the contribution of the triangle singularity later.}
For the other two $\Xi^*$ states, their quantum numbers are not determined yet, and~we will consider various $J^P$ possibilities~later.

\begin{figure}[h]
\centering
	\includegraphics[width=0.95\linewidth]{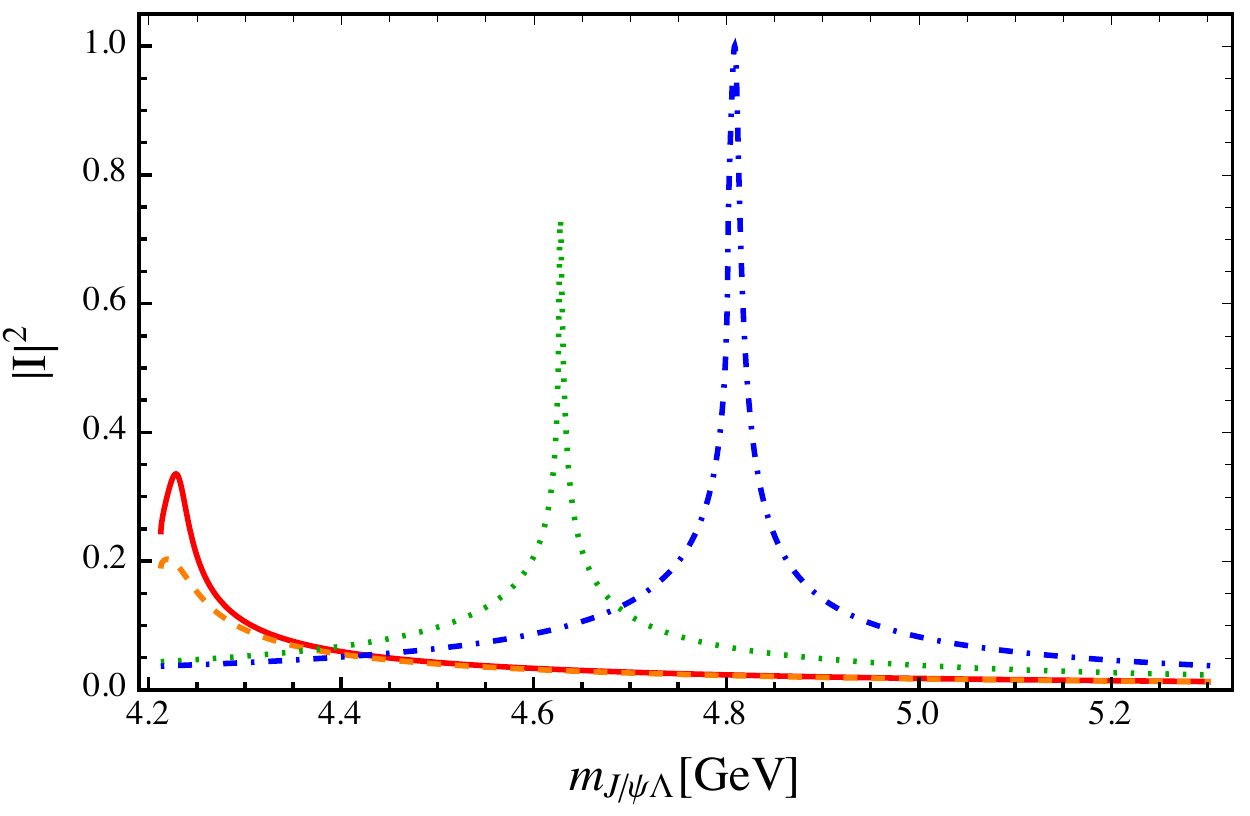}
	\caption{The value of $|I|^2$ in the $m_{J/\psi \Lambda}$ invariant mass distribution. The~red solid (orange dashed), green dotted and blue dotdashed lines correspond to the Nos.~1, 4 and 11 cases in Table~\ref{table:TS}, respectively. For~the case of No.1, the~red solid and orange dashed lines represent the different choices of the $\Xi(2500)$ width: the narrower and wider peaks correspond to taking 59 and 150~MeV for the width, respectively.
	}
  \label{Fig:4TS}
\end{figure}

The $m_{J/\psi \Lambda}$ distribution for $|I|^2$ with an arbitrary normalization factor of these three cases is shown in Figure~\ref{Fig:4TS}.
For the widths of the involved charmonium, we quote their central values given in the RPP~\cite{Zyla:2020}.
While for some of the $\Xi^*$ states like the $\Xi(2120)$ and $\Xi(2500)$, the~widths are not precisely determined.
For the width of the former, we take the upper bound of the given range, which is $20$ MeV.
In the latter case, two values were reported by different experimental groups, which are $150$~\cite{Alitti:1969rb} and $59$~MeV~\cite{Bartsch:1969qs}.
We give results using these two values.
As shown in Figure~\ref{Fig:4TS}, the~triangle singularity from the $J/\psi \Xi(2500) \Lambda$ loop diagram is much broader than the other two cases.
As discussed above, the~triangle loop diagram that involves the $\Xi(2030) \bar{K} \Lambda$ vertex should be severely suppressed because of the existence of the $D$- or higher wave interaction,
while for the $\Xi(2120)$ involved vertices, there~could be lower partial-wave interactions.
The contribution from the $\Xi(2120)\chi_{c1}\Lambda$ loop (the blue dotdashed line in Figure~\ref{Fig:4TS}) is expected to be much larger than that from the $\Xi(2030)\psi(2S)\Lambda$ (the green dotted line) once all the interactions at the vertices are~included.

Combining all the analysis above, the~$\Xi(2120)\chi_{c1}\Lambda$ loop is found to be the most promising process to generate an observable triangle singularity in the $\Xi^-_{b} \to K^- J/\psi \Lambda$ decay ({We should also note that the $\Xi(2120)$ is just a one-star state~\cite{Zyla:2020}, and~its existence needs further confirmation}).
We will calculate and discuss the structures from this triangle loop diagram in detail in the next~section.

\section{Detailed Analysis of the Amplitudes for the Diagram with \texorpdfstring{$\bm{\chi_{c1} \Xi(2120) \Lambda}$}{chic1 Xi Lambda} Loop} \label{sec:analysis}

We will focus on the triangle diagram involving the $\chi_{c1}$ and the $\Xi(2120)$ in this section, as~it is the most promising diagram to have a sizeable triangle singularity contribution.
The specific interactions of vertices are taken into account in this section, here we are going to study them by constructing effective Lagrangians with the lowest number of derivatives.
At present, the~quantum numbers of the $\Xi(2120)$ state and the $J/\psi \Lambda$ system are both flexible, and~the former affects the $\Xi_b \Xi^* \chi_{c1}$ and $\Xi^* \Lambda K^-$ vertices, while the later influences the $\chi_{c1} \Lambda J/\psi \Lambda$ vertex.
Let us discuss them one by~one.

Firstly, let us consider different choices of the spin-parity quantum numbers of the $\Xi(2120)$, which~are still unknown.
The two vertices involving the $\Xi(2120)$ in this diagram are $\Xi_b \Xi(2120) \chi_{c1}$ and $\Xi(2120) \Lambda \bar K$.
In Table~\ref{table:PW}, the~assumed spin-parity of $\Xi(2120)$ and the corresponding lowest allowed partial waves of these two vertices are listed.
{In the Lorentz covariant orbital-spin coupling scheme, the~interaction of a vertex is directly related to its partial wave $L$ and the relative momentum of the two final particles, i.e.,~the strength of interaction is proportional to $q^L$. Therefore, the~$D$- or higher partial waves are neglected since the momentum is very small around where the triangle singularity happens.}
%
Thus, $J^P=1/2^+$, $1/2^-$ and $3/2^+$ possibilities are not suppressed by high partial waves. Among~these three cases, the~$J^P=1/2^-$ possibility is singled out as both vertices are in $S$ waves, and~the strength of the amplitude in this case should be stronger than that of the other cases.
Therefore, we will calculate the amplitude with $J^P(\Xi^*)=1/2^-$ to check the structure behavior from the triangle~singularity.

\begin{table}[tb]
	\caption{\label{table:PW}The interactions of the $\Xi_b \Xi(2120) \chi_{c1}$ and $\Xi(2120) \Lambda \bar K$ vertices with different spin-parities of $\Xi(2120)$.}
	\renewcommand\arraystretch{1.5}
	\begin{tabular}{p{2.0cm}<{\centering}p{3.0cm}<{\centering}p{3.0cm}<{\centering}}
		\hline
		$J^P$ & $\Xi_b \Xi(2120) \chi_{c1}$  & $\Xi(2120) \Lambda \bar K$ \\
		\hline
		$1/2^+$	& $S$-wave & $P$-wave	\\
		$1/2^-$	& $S$-wave & $S$-wave	\\
		$3/2^+$ & $S$-wave & $P$-wave	\\
		$3/2^-$ & $S$-wave & $D$-wave	\\
		$5/2^+$ & $P$-wave & $F$-wave	\\
		$5/2^-$ & $P$-wave & $D$-wave	\\
		\hline
	\end{tabular}
\end{table}

We then consider the quantum numbers of the $J/\psi \Lambda$ in the final state for fixed spin-parity of the $\Xi(2120)$.
For the same reason mentioned before, the~$D$- or higher partial wave interactions are not considered in the present discussion.
If the quantum numbers of the $J/\psi \Lambda$ system are $1/2^+$ or $3/2^+$, the~$\chi_{c1} \Lambda$ system can be in an $S$-wave and the $J/\psi \Lambda$ in a $P$-wave.
On the contrary, if~the quantum numbers of the $J/\psi \Lambda$ system are $1/2^-$ or $3/2^-$, there is a $P$-wave in the $\chi_{c1} \Lambda$ system and an $S$-wave in the $J/\psi \Lambda$ system.
In both cases, there are an $S$-wave and a $P$-wave for this four-particle-vertex, but~the corresponding behaviors of these two triangle loop integrations are different.
{
As discussed in the previous paragraph, the~effective Lagrangian of a $P$-wave interaction is proportional to the relative momentum of the involved particle.
Meanwhile, at~the singularity point, the~invariant mass $m_{J/\psi \Lambda}$ is not far away from the $\chi_{c1}\Lambda$ threshold, and~both the $\chi_{c1}$ and the $\Lambda$ are on-mass-shell. It tells that the momenta of the $\chi_{c1}$ and $\Lambda$ are small.
Thus, for~the negative parity case,  the~$P$-wave is in the $\chi_{c1}\Lambda$ system, which provides a loop momentum $q$ to the loop integral and suppresses the singular behavior around the $\chi_{c1}\Lambda$ threshold, while the $P$-wave interaction of the $J/\psi \Lambda$ system for the positive parity case is independent of the loop integration.}
%
In the next step, we will take the quantum numbers of the $J/\psi \Lambda$ system as $1/2^+$ and $1/2^-$, respectively, to~investigate how the finial results are~affected.

We first take the quantum numbers of the $J/\psi \Lambda$ system being $1/2^+$ and the spin-parity of the $\Xi(2120)$ are assumed to be $1/2^-$.
The amplitudes for the vertices are constructed as
\begin{eqnarray}
	t_{\Xi_b \Xi^* \chi_{c1}} &=& g_a \bar{u}(P-q) \gamma^\mu u(P) \epsilon^*_\mu(q),	\nonumber \\
	t_{\Xi^* \Lambda \bar K} &=& g_b \bar{u}(k_1-q) u(P-q), \nonumber \\
	t_{\Lambda \chi_{c1} \Lambda J/\psi} &=& g_c \epsilon_\mu(q) \bar{u}(k_1-k_2) ({\Slash k_1}+\sqrt{k_1^2}) \gamma_5 (\gamma^\mu-\frac{k_1^\mu {\Slash k_1}}{k_1^2}) \nonumber \\
	&& \times u(k_1-q) \epsilon_\nu^*(k_2) k_1^\nu \frac{k_1^2+m_{J/\psi}^2-m_{\Lambda}^2}{k_1^2},
\end{eqnarray}
here we follow the momentum conventions given in Figure~\ref{Fig:feyndiag4}a.
The amplitude of this triangle diagram~reads
\begin{eqnarray}
	\mathcal{M} &=& g	\int \frac{d^4 q}{(2\pi)^4} \bar{u}(k_1-k_2) ({\Slash k_1}+\sqrt{k_1^2}) \gamma_5 (\gamma^\mu-\frac{k_1^\mu {\Slash k_1}}{k_1^2}) \nonumber\\
	&& \times \frac{{\Slash k_1}-{\Slash q}+m_3}{(k_1-q)^2-m_3^2} \frac{{\Slash P}-{\Slash q}+m_2}{(P-q)^2-m_2^2+i m_2 \Gamma_2} \nonumber\\
	&& \times \frac{-\gamma_{\mu}+\frac{q_\mu {\Slash q}}{m_1^2}}{q^2-m_1^2+i m_1 \Gamma_1} u(P) \frac{k_1^2+m_{J/\psi}^2-m_{\Lambda}^2}{k_1^2} \nonumber\\
	&& \times \epsilon_\nu^*(k_2) k_1^\nu.
\end{eqnarray}

Then we calculate the decay with the quantum numbers of the $J/\psi \Lambda$ system being $1/2^-$.
The~amplitude for the $\Lambda \chi_{c1} \Lambda J/\psi$ vertex reads
\begin{eqnarray}
	 t_{\Lambda \chi_{c1} \Lambda J/\psi}&=&g_c \bar{u}(k_1-k_2) \gamma_5 (\gamma^\mu-\frac{k_1^\mu {\Slash k_1}}{k_1^2}) ({\Slash k_1}+\sqrt{k_1^2}) \nonumber\\
	&& \times u(k_1-q) \epsilon_\mu^*(k_2) \epsilon_\nu(q) k_1^\nu \frac{2k_1 \cdot q}{k_1^2},
\end{eqnarray}
and the amplitude in this case is
\begin{eqnarray}
	\mathcal{M} &=& g	\int \frac{d^4 q}{(2\pi)^4} \bar{u}(k_1-k_2) \gamma_5 (\gamma^\mu-\frac{k_1^\mu {\Slash k_1}}{k_1^2}) ({\Slash k_1}+\sqrt{k_1^2}) \nonumber\\
	&& \times \frac{{\Slash k_1}-{\Slash q}+m_3}{(k_1-q)^2-m_3^2} \frac{{\Slash P}-{\Slash q}+m_2}{(P-q)^2-m_2^2+i m_2 \Gamma_2}  \nonumber\\
	&& \times \frac{-{\Slash k_1}+\frac{\Slash q}{m_1^2} k_1 \cdot q}{q^2-m_1^2+i m_1 \Gamma_1} u(P) \frac{2k_1 \cdot q}{k_1^2} \epsilon_\mu^*(k_2).
	\label{eq:m12m}
\end{eqnarray}

It should be noticed that the absolute values of the coupling constants in both cases are unknown.
Therefore, our results are normalized to 1 at the maximum.
The differential decay width in the rest frame of the initial particle $\Xi_b$ is
\begin{equation}
{\mathrm{d} \Gamma = \frac1{(2\pi)^3} \frac1{32m_{\Xi_b}^3} \overline{{\left| \mathcal M \right|}^2} \mathrm{d}m_{K \Lambda}^2 \mathrm{d}m_{J/\psi \Lambda}^2.}
	\label{eq:dGamma}
\end{equation}

\begin{figure}[h]
\centering
\subfigure []{\label{Fig:dGamma1}
	\includegraphics[width=0.95\linewidth]{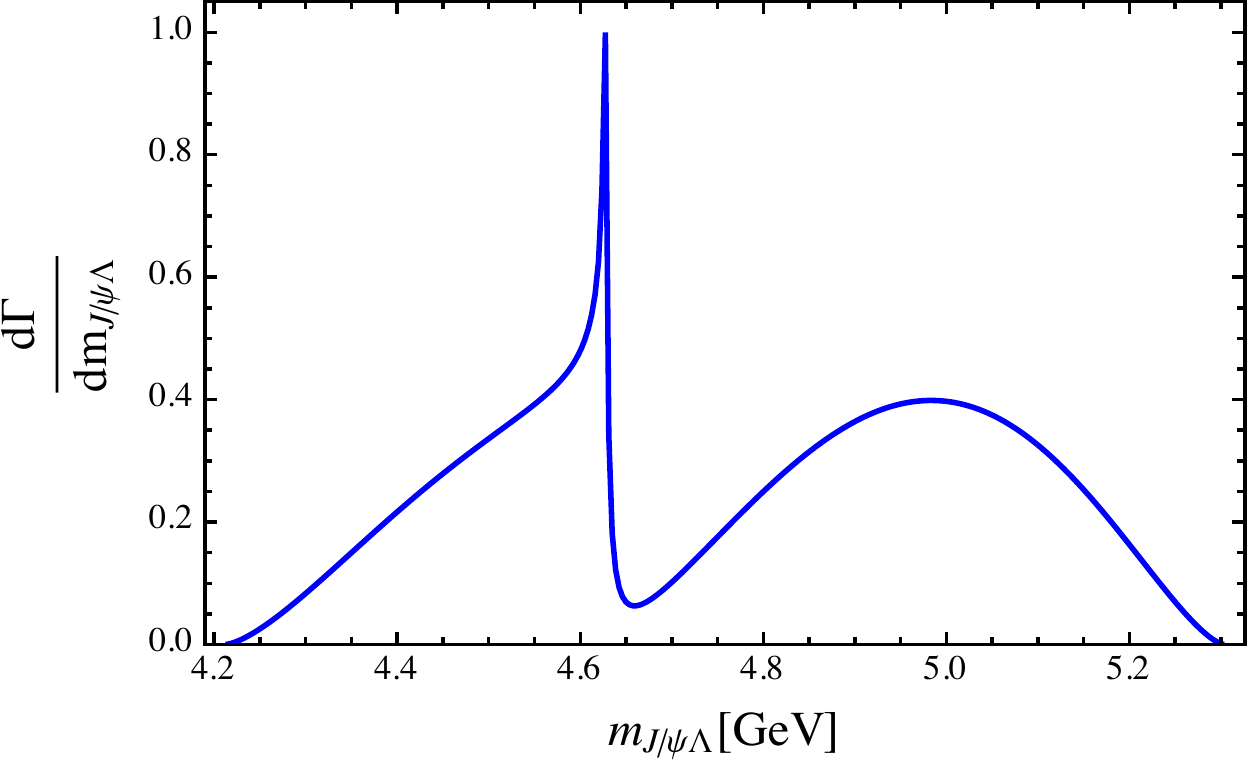}} \\ 
\subfigure[] {\label{Fig:dGamma2}
	\includegraphics[width=0.95\linewidth]{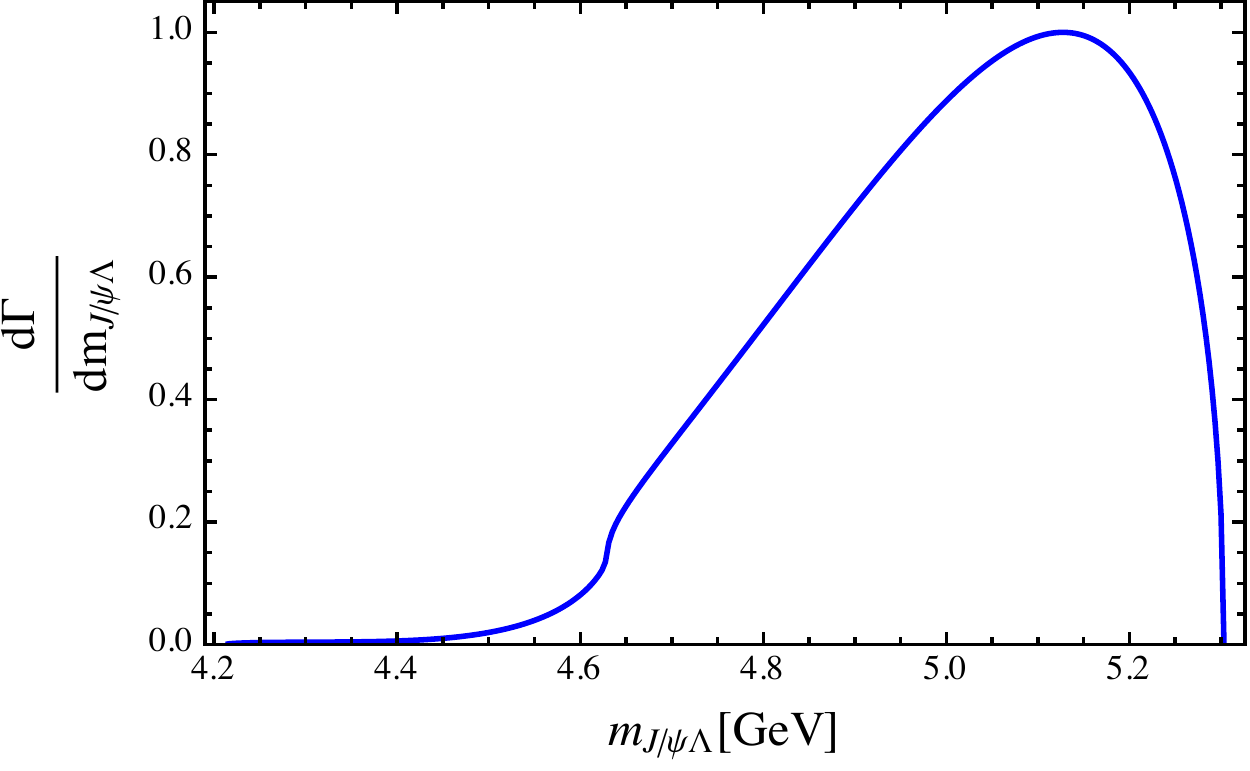}}
\caption{The $m_{J/\psi \Lambda}$ invariant mass distribution for the $\Xi^-_{b} \to K^- J/\psi \Lambda$ process via the $\chi_{c1} \Xi(2120) \Lambda$ loop with the quantum numbers of the $J/\psi \Lambda$ system being (\textbf{a}) $1/2^+$, and~(\textbf{b}) $1/2^-$, respectively.
  \label{Fig:dGamma}}
\end{figure}

{After integrating $\mathrm{d}m_{K \Lambda}^2$, t}he normalized differential decay widths in the $J/\psi \Lambda$ invariant mass distribution in these two cases are shown in Figure~\ref{Fig:dGamma} with the widths of the exchanged particles taken into~consideration.

It can be seen that in the $1/2^+$ case the peak structure from the triangle singularity emerges at $m_{J/\psi \Lambda}=4.628$~GeV and is distinct enough comparing to the background.
However, this structure could not be observed any more in the $1/2^-$ case.
Naively one would expect that the peak should still be there since the numerator of the integral would not diminish the singularity generated from the denominator.
Then here arises a question why the $P$-wave $\chi_{c1} \Lambda$ in the four-particle vertex can suppress the singular behavior so much.
When constructing the effective Lagrangian of the four-particle vertex in the Lorentz covariant orbital-spin coupling scheme~\cite{Zou:2002yy}, the~$P$-wave interaction in the $\chi_{c1} \Lambda$ system introduces a factor of the momentum difference of the exchanged $\Lambda$ and $\chi_{c1}$.
Thus, the~amplitude is directly proportional to the velocity difference between these two particles in the $1/2^-$ case.
As~discussed above, when the triangle singularity happens, all the three intermediate particles are on the mass shell and the process of each vertex actually happens classically.
In this $\chi_{c1} \Xi(2120) \Lambda$ loop diagram, we~find that the velocity of the intermediate $\Lambda$ is very close to that of the $\chi_{c1}$ when the singularity occurs.
In~other words, their relative velocity is quite small in the rest frame of $\Xi_b$, meaning that the singularity is located very close to the $\chi_{c1}\Lambda$ threshold.
More specifically, the~factor $\left(-{\Slash k_1}+\frac{\Slash q}{m_1^2} k_1 \cdot q\right)$ in Equation~\eqref{eq:m12m} from the $P$-wave $\chi_{c1} \Lambda$ system in the triangle singularity region can be derived as~follows,
\begin{equation}
 -{\Slash k_1}+\frac{\Slash q}{m_1^2} k_1 \cdot q \sim \frac{|\vec{q}|}{q^0}-\frac{|\vec{k}_1|}{k_1^0} \sim v_{\Lambda}-v_{\chi_{c1}}\sim 0,
\end{equation}
explaining the suppression of the singular behavior in this~case.

Since the amplitude is actually related to the partial wave interactions of the $\chi_{c1} \Lambda \to J/\psi \Lambda$ vertex, the~$3/2^+$ and  $3/2^-$ cases should be similar to the $1/2^+$ and $1/2^-$ cases, respectively.
This~indicates that this $\Xi(2120) \chi_{c1} \Lambda$ loop diagram could generate a peak in the $m_{J/\psi \Lambda}$ mass spectrum for the $\Xi^-_{b} \to K^- J/\psi \Lambda$ decay when the quantum numbers of the $J/\psi \Lambda$ system are $1/2^+$ or $3/2^+$.
We~then note that the threshold of the $\chi_{c1}\Lambda$ is at 4.626~GeV, which gives rise to a cusp (for the $S$-wave) in the distribution of the differential decay width.
As the positions of the threshold and the triangle singularity (at about 4.628~GeV) are quite close, these two structures would lead to only one peak around this energy as shown in Figure~\ref{Fig:dGamma}a.
Thus, if~a structure is observed in this region in future experiments, the~triangle singularity effects need to be taken into account for the $J^P=\frac12^+$ or $\frac32^+$ quantum~numbers.

There are a lot of similarities between the $\Lambda_b \to K^- J/\psi p$ and $\Xi_b \to K^- J/\psi \Lambda$ through triangle loop diagrams.
The $\Lambda_b \Lambda^* \chi_{c1}$ vertex, as~considered in Refs.~\cite{Guo:2015umn,Bayar:2016ftu}, and~the $\Xi_b \Xi^* \chi_{c1}$ vertex involve the same product of CKM matrix elements and should share a similar coupling.
The strength of the two four-particle-vertices $\chi_{c1} p \to J/\psi p$ and $\chi_{c1} \Lambda \to J/\psi \Lambda$ should also be similar due to the approximate light-flavor SU(3) symmetry.
For the two rest vertices $\Lambda(1890) K^- p$ and $\Xi(2120) K^- \Lambda$, the~$\bar K N$ channel occupies a rate of 20--35\% among all the decay modes of $\Lambda(1890)$, while the branching ratio of $\Xi(2120)$ is still unknown and $\bar K \Lambda$ is the only observed decay channel.
We can roughly regard the strengths of these two vertices to be of the same order of magnitude.
Thus, studying possible structures in the  $\Xi_b \to K^- J/\psi \Lambda$ decay around 4.63~GeV should also provide useful insights into the possible role of triangle singularity in the $\Lambda_b \to K^- J/\psi p$ decay around 4.45~GeV.
Yet, the~existence of  one-star $\Xi(2120)$ state needs to be confirmed, which can be studied by analyzing the $\Lambda \bar K$ invariant mass distribution of the process under~discussion.

\section{Summary} \label{sec:summary}

Inspired by the analysis of the triangle singularities in the $\Lambda_b \to K^- J/\psi p$ process in Ref.~\cite{Bayar:2016ftu}, we~apply the same method to the $\Xi_b \to K^- J/\psi \Lambda$ process via charmonium-$\Xi^*$-$\Lambda$ intermediate states to check whether there exist possible triangle singularities.
We find that there are 11 and 7 possible triangle singularities with various combinations of charmonium-$\Xi^*$ states and $D_S^*$-$\Xi_c^*$ states, respectively.
For~the former, the~$\chi_{c1} \Xi(2120) \Lambda$ loop diagram is expected to be the most promising process to observe the singularity structure; for the latter, those with $\bar D_{s1}^*(2700)$ listed in Table~\ref{table:TSbc} should be more important than the others. Nevertheless, the~absolute values of all these amplitudes could not be precisely evaluated without knowing the amplitudes for all the involved vertices.
For the $\chi_{c1} \Xi(2120) \Lambda$ diagram,
when the spin-parity quantum numbers of the $J/\psi\Lambda$ system are $1/2^+$ or $3/2^+$, the~joint effects of the triangle singularity and the $\chi_{c1}\Lambda$ threshold would produce a narrow peak around 4.63~GeV in the $J/\psi\Lambda$ invariant mass distribution.
A study of such effects needs to be taken into account in the search of hidden-charm strange pentaquarks, and~a study of the $\Xi_b \to K^- J/\psi \Lambda$ will also shed light on the role of triangle singularities in the $\Lambda_b\to K^- J/\psi p$ process.

\section*{Acknowledgments}

We are grateful to Eulogio Oset for a careful reading of our~manuscript. This work is supported in part by the Fundamental Research Funds for the Central Universities,  by~the National Natural Science Foundation of China (NSFC) under Grants No.~11835015, No.~11947302 and No.~11961141012, by~the Chinese Academy of Sciences (CAS) under Grants  No.~XDB34030303 and No.~QYZDB-SSW-SYS013, by~NSFC and the Deutsche Forschungsgemeinschaft (DFG) through the funds provided to the Sino-German Collaborative Research Center ``Symmetries and the Emergence of Structure in QCD'' (NSFC~Grant No. 11621131001, DFG Grant No. CRC110), and~by the CAS Center for Excellence in Particle Physics~(CCEPP).

\bibliography{main}

\end{document}